\documentclass[%
 reprint,
 nofootinbib,
 amsmath,amssymb,
pra,
floatfix,
superscriptaddress
]{revtex4-2}
\usepackage[utf8]{inputenc}

\usepackage{amssymb}
\usepackage{mathtools}
\usepackage{siunitx}
\usepackage{physics}
\usepackage{hyperref}
\usepackage{graphicx}

\begin{document}

\title{Light-Pulse Atom Interferometric Test of Continuous Spontaneous Localization}

\author{Sascha Vowe}
 \affiliation{
 Institut für Physik, 
 Humboldt-Universit\"{a}t zu Berlin, 
 Newtonstra{\ss}e 15, 12489 Berlin, Germany
}

\author{Sandro Donadi}
 \email{sandro.donadi@ts.infn.it}
 \affiliation{%
 Istituto Nazionale di Fisica Nucleare, Trieste Section, Via Valerio 2, 34127, Trieste, Italy
}

\author{Vladimir Schkolnik}
 \affiliation{
 Institut für Physik, 
 Humboldt-Universit\"{a}t zu Berlin, 
 Newtonstra{\ss}e 15, 12489 Berlin, Germany
}

\author{Achim Peters}
\affiliation {
Institut für Physik, 
 Humboldt-Universit\"{a}t zu Berlin, 
 Newtonstra{\ss}e 15, 12489 Berlin, Germany
}

\author{Bastian Leykauf}
\email{leykauf@physik.hu-berlin.de}
 \affiliation{
 Institut für Physik, 
 Humboldt-Universit\"{a}t zu Berlin, 
 Newtonstra{\ss}e 15, 12489 Berlin, Germany
}

\author{Markus Krutzik}%
\affiliation{
Institut für Physik, 
 Humboldt-Universit\"{a}t zu Berlin, 
 Newtonstra{\ss}e 15, 12489 Berlin, Germany
}

\date{\today}

\begin{abstract}
We investigate the effect of the Continuous Spontaneous Localization (CSL) model on light-pulse atom interferometry. Using a path-integral approach with an additional stochastic potential accounting for CSL, we derive an exponential loss of the contrast that scales linearly with the interferometer time $T$ if both interferometer arms are spatially separated. We compare our theoretical results with measurements from a cold rubidium atom interferometer based on counter-propagating two-photon transitions with pulse separation times up to $T = \SI{260}{ms}$ and obtain the corresponding bounds on the CSL parameters.

\end{abstract}

\maketitle

\section{Introduction}

Collapse models are popular in giving a realistic interpretation of the wave function collapse, which solves the measurement problem. In these models, the Schr\"{o}dinger equation is extended by adding terms that describe the collapse of the superposition of a particle's wave function in position space into a localized state \cite{GRW1986, bassi2003dynamical, bassi2013models}. In the Continuous Spontaneous Localization (CSL) model \cite{CSL1990}, the non-linear and stochastic terms describing the collapse depend on two newly introduced phenomenological parameters $\lambda_\text{CSL}$ and $r_C$. $\lambda_\text{CSL}$ is the collapse rate and sets the strength of the collapse, while $r_C$ is a length which sets the spatial resolution of the collapse. Bounds on these parameters have been deduced for various experiments and physical systems \cite{Bilardello2016, carlesso2016experimental, piscicchia2017csl, vinante2017improved,Fein2019,PhysRevD.99.103001,PhysRevLett.125.100404,PhysRevResearch.2.013057}, for a recent review see \cite{Carlesso2019, carlesso2022present}.

The best interferometric tests of the CSL model currently stem from Talbot-Lau interferometry using molecules that have a mass of $\sim \SI{25 000}{u}$ \cite{Fein2019}, setting the bound of $\lambda_\text{CSL} \lesssim \SI{e-7}{\per\second}$ for $r_{C} = \SI{e-7}{m}$.

In this paper we study CSL effects on light-pulse cold atom interferometry, more specifically on the contrast of the measured interference fringes. Similar work has been reported in \cite{Schrinski2017} with a focus on many-body enhanced decoherence based on interferometry with Bose-Einstein condensates.

Our theoretical analysis is simplified compared to solving the CSL nonlinear equation, since we are only interested in expectation values. Then, the non-linear CSL dynamics can be replaced by a stochastic Schr\"{o}dinger equation \cite{Adler2007}
\begin{equation}
    i\hbar\frac{\mathrm{d}}{\mathrm{d}t}|\psi_t\rangle = \left(\hat{H}_0 + \hat{V}_\text{CSL} \right)|\psi_t\rangle, \label{eq:CSL-schrodinger-equation}
\end{equation}
where $\hat{V}_\text{CSL}$ is an additional stochastic potential which accounts for CSL. Eq. (\ref{eq:CSL-schrodinger-equation}) does not lead to wave function collapse, yet it leads to the same master equation and, therefore, the same expectation values as the CSL non-linear equation. 
For composite systems, the CSL effective stochastic potential takes the form

\begin{equation}\label{V_csl}
\begin{multlined}
V_\text{CSL}(\hat{\vb{q}}_1\cdots, \hat{\vb{q}}_N,t) \\
=-\frac{\hbar\sqrt{\lambda_\text{CSL}}}{m_0(\sqrt{\pi}r_{C})^{3/2}}\int \mathrm{d}\vb{r}\,w(\vb{r},t)\sum_{k=1}^{N}m_ke^{-\frac{(\vb{r}-\hat{\vb{q}}_{k})^{2}}{2r_{C}^{2}}},
\end{multlined}
\end{equation}
where $N$ is the number of the nucleons composing the system, $\hat{\vb{q}_k}$ their positions, $m_0$ is one atomic mass unit and $w(\vb{r},t)$ is Gaussian white noise that is characterized by the expectation values $\mathbb{E}[w(\vb{r},t)] = 0$ and correlations $\mathbb{E}[w(\vb{r},t) w(\vb{r}',t')] = \delta(\vb{r}-\vb{r}') \delta (t-t')$.
The electrons' contribution is omitted because CSL is mass proportional, and their masses are negligible compared to that of the nucleons.

In the case of an atom, we can simplify eq. (\ref{V_csl})  further by approximating the protons' and neutrons' positions (up to a difference of order of femtometer, much smaller than the range of possible $r_C$) $\hat{\vb{q}}_{k}=\hat{\vb{q}}$, where $\hat{\vb{q}}$ is the position operator of the center of mass of the nucleus. We will also not consider the small difference in mass between neutrons and proton, hence they all have the same mass $m=m_{0}$. 
In this case, we can write eq. (\ref{V_csl}) as
\begin{equation}
\begin{multlined}
    V_\text{CSL}(\hat{\vb{q}},t) = -\frac{\hbar \sqrt{\lambda_\text{CSL} }N}{(\sqrt{\pi}r_C)^{3/2}}\\
    \times \int \mathrm{d}\vb{r}\; w(\vb{r},t) \exp (-\frac{(\vb{r}-\hat{\vb{q}})^2}{2 r_C^2}).
\end{multlined}
\end{equation}
This additional potential alters the time evolution of atomic systems that can be observed in precision measurements using atom interferometry.

In this paper, we employ cold atom interferometry in a fountain configuration to test CSL. By evaluating the contrast for different interrogation times in a symmetric Mach-Zehnder configuration, we can place upper bounds on the collapse rate $\lambda_\text{CSL}$ for different values of $r_C$. Towards this, we first derive the contrast of the interference fringes in a Mach-Zehnder interferometer (MZI) using a path-integral approach. Afterwards, we give a brief description of our light-pulse atom-interferometer experiment and present the experimental sequence and data used to put bounds on $\lambda_\text{CSL}$ for different values of $r_C$. 
In the conclusion, we discuss limitations and prospective developments in light-pulse atom interferometry that could improve this bound.

\begin{figure}
    \centering
\begin{minipage}{0.48\textwidth}
\includegraphics[width=0.9\textwidth]{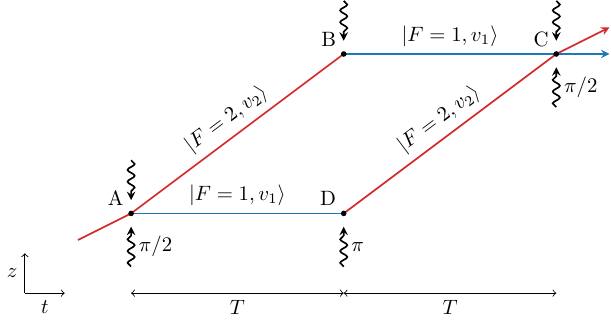}
\end{minipage}
    \caption{Atom interferometer in Mach-Zehnder configuration using stimulated Raman transitions between two hyperfine ground states $\ket{F=1}$ and $\ket{F=2}$. The paths are used in the path integral approach outlined in the text.}
    \label{fig:ai-sketch}
\end{figure}{}

\section{Contrast loss mechanism in atom interferometry induced by CSL}\label{sec:theory}

Light-pulse atom interferometry is an important tool in modern metrology and has been used to measure inertial effects in e.g., gravitational accelerometers \cite{Peters2001}, gravity gradiometers \cite{McGuirk2002} and gyroscopes \cite{Gustavson1997}. It also found many applications in fundamental physics through measurements of the fine-structure constant \cite{Wicht2002,Yu2019},
the gravitational constant \cite{Bertoldi2006},
tests of Lorentz invariance \cite{Chung2009},
general relativity \cite{Dimopoulos2007}, dark-energy theories \cite{Hamilton2015}, or the universality of free fall \cite{Schlippert2014}.

In this work, we present a light-pulse atom interferometric test of CSL. In order to compute the phase difference $\Delta \phi_\text{CSL}$ due to the presence of CSL, we follow the perturbative path integral approach described in \cite{Tannoudji1994}, for a Mach-Zehnder configuration (see fig. \ref{fig:ai-sketch}):

\begin{equation}\label{eq.phi_CSL}
\Delta\phi_\text{CSL} =\frac{1}{\hbar}\ointop_{ABCDA}V_\text{CSL}(z(t), t)\mathrm{d}t.
\end{equation}
We note that, precisely speaking, the CSL effective potential depends on the position $\vb{q}(t)$ in all the three space directions. However, for the effect we are computing, only the coordinate along the $z$ direction is relevant (see section \ref{app:a} of the appendix).

Since $V_\text{CSL}$ is not deterministic, all predictions need to be computed by averaging over the noise. However, the averaged phase difference acquired during passage of the atom interferometer vanishes as a consequence of $\mathbb{E}[w(\vb{r},t)]=0$. 
Therefore, CSL effects can not be observed as a phase shift.

On the other hand, the noise has a nonzero second moment since $\mathbb{E}[w(\vb{r},t) w(\vb{r}',t')] = \delta(\vb{r}-\vb{r}') \delta (t-t')$.
Therefore, we expect deviations in the variance of $\Delta\phi_\text{CSL}$.
Through a straightforward calculation reported in section \ref{app:a} of the appendix we show that
\begin{widetext}
\begin{equation}\label{eq:phivariance}
    \mathbb{E}[\Delta\phi^2_\text{CSL}] =
    \mathbb{E}\bigg[\frac{1}{\hbar^2} \ointop_{ABCDA}\ointop_{ABCDA}
    V_\text{CSL}(z(t), t)V_\text{CSL}(z(t'), t')\mathrm{d}t\mathrm{d}t'\bigg]=
    4 \lambda_\text{CSL} N^2 T \Bigg[
    1 - \frac{\sqrt{\pi}}{2}\frac{\erf \Big(\frac{v_2-v_1}{2 r_C}T\Big)}{\frac{v_2-v_1}{2 r_C}T}
    \Bigg]. 
\end{equation}{}
\end{widetext}
Here, $v_1$ and $v_2$ are the velocities of the two arms of the atom interferometer as shown in fig. \ref{fig:ai-sketch}. In the case of our light-pulse atom interferometer, this velocity difference is induced by driving Raman transitions with an effective wave number $k_{\text{eff}} = (v_2 - v_1) / (\hbar m_\text{Rb})$ and $m_\text{Rb}$ the mass of a $^{87}$Rb atom.

A variance in the expectation value of the phase will result in a loss of fringe contrast of the atom interferometer which is determined by the relative population of the two ground states.
This relative state population depends on the relative phase difference and is described by
\begin{equation}\label{eq:coherence}
    P = \frac{1}{4} \mathbb{E} \Big[\big|1 + \exp (i \Delta\phi_0)  \cdot \exp (i \Delta\phi_\text{CSL}) \big|^2 \Big]
\end{equation}
where the phase difference in a light-pulse MZI is~\cite{Antoine2006,Cheinet2006}
\begin{equation}\label{phi0}
    \Delta\phi_0 =
    \Big(k_{\text{eff}} g - \alpha \Big) T^2,
\end{equation}
with the gravitational acceleration $g$ (we omit a gravity gradient), $\alpha$ the two-photon detuning chirp rate and $T$ the pulse separation time of the MZI. By choosing an appropriate chirp rate $\alpha$, it is possible to compensate for the phase shift induced due to gravity by effectively moving to the frame falling with the atoms \cite{Young1997}, i.e. the frame depicted in fig. \ref{fig:ai-sketch}. In this case, the phase difference $\Delta\phi_0$ in eq. \eqref{phi0} is zero for all $T$. 

The average over the noise in eq. \eqref{eq:coherence} can be computed exactly since the noise is Gaussian to obtain
\begin{equation}\label{eq:fringes_alpha_scan}
     P=\frac{1}{2} \Bigg[1 + \exp (- \frac{\mathbb{E}[\Delta\phi^2_\text{CSL}]}{2})
    \cdot \cos (\Delta\phi_0 )
    \Bigg]\text{.}
\end{equation}
Our result shows an exponential loss of contrast. The strength of this loss depends on the collapse rate $\lambda_\text{CSL}$, the correlation length $r_C$, the number of nucleons squared $N^2$, the time of the experiment $2T$ and the velocities of the two wave packets $v_1$ and $v_2$.

According to eqs. (\ref{eq:phivariance}) and (\ref{eq:fringes_alpha_scan}), there are two relevant regimes: (i)~When $r_C$ is much larger than the maximum distance between the interferometer arms $(v_2-v_1)T\sim \SI{e-3}{m}$, the loss of contrast gets smaller as $r_C$ increases. Consequently, the bound on $\lambda_\text{CSL}$ gets weaker for larger $r_C$.
(ii)~When $r_C\ll (v_2-v_1)T\sim \SI{e-3}{m}$, the error function in eq. (\ref{eq:phivariance}) becomes negligible and the damping factor $\mathbb{E}[\Delta\phi^2_\text{CSL}]$ becomes independent of $r_C$. Therefore, the bound on $\lambda_\text{CSL}$ becomes constant, i.e. independent of $r_C$ as well.

While having the merit of being relatively simple and giving a good insight into the physics of the problem, the analysis presented until now completely neglects the finite size of the atomic wave packets $\ell$. This approach is a reasonable approximation as far as $r_C\geq \ell$. However, for smaller $r_C$ a wave packet analysis is required. This analysis is detailed in section \ref{app:b} of the appendix : In order to observe interference, we require that the two wave packets in the interferometer have sufficient spatial overlap at time $2T$. This requirement leads to an upper bound for the ratio $\lambda_\text{CSL}/r_C^2$, i.e. the bound gets stronger as $r_C$ decreases. For the experimental parameters investigated in this work,
this bound is stronger than the one obtained from eqs. (\ref{eq:phivariance}) and (\ref{eq:fringes_alpha_scan}) (which neglects the finite wave packet size) if $r_C\leq \SI{3.8e-6}{m}$.

\section{Experiment \& Results}\label{sec:exp}

\begin{figure}
    \centering
\begin{minipage}{0.48\textwidth}
\includegraphics[width=0.55\textwidth]{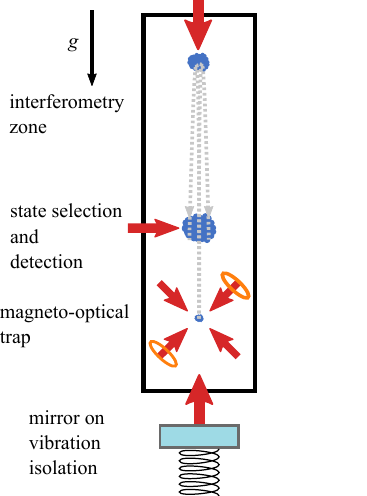}
\end{minipage}
    \caption{Sketch of experimental apparatus, showing its functional subsections, the trajectory of the atomic cloud as well as optical access to the vacuum chamber indicated by arrows. For more details on the experimental sequence, see the main text.}
    \label{fig:gain-sketch}
\end{figure}

In this section we describe the measurement protocol for multiple contrast measurements over different interferometry times $2T$. We use the observed loss in contrast to estimate an upper bound for $\lambda_\text{CSL}$ in equation \eqref{eq:fringes_alpha_scan}.
See fig.\,\ref{fig:gain-sketch} for a sketch and \cite{Hauth2013} for a more detailed description of the experimental apparatus.

The atom interferometer is formed by driving stimulated Raman transitions between the two hyperfine ground states $F=1$ and $F=2$ of $^{87}$Rb by counter-propagating laser beams. The velocity difference between the two arms of interferometer is $v_1-v_2 \approx \SI{11e-3}{\meter\per\second}$. For a single measurement, a laser-cooled cloud of about $\num{e9}$ $^{87}$Rb atoms is launched upwards using moving molasses with a temperature of $\sim\SI{2}{\micro\kelvin}$ into a magnetically shielded interferometry tube. Atoms within a narrow velocity class along the $z$ direction corresponding to a vertical temperature of $\sim\SI{80}{nK}$ are selected by Doppler sensitive stimulated Raman transitions and the other atoms are removed from the sample by a blow-away laser pulse. After selecting atoms in the first-order magnetically insensitive $m_F=0$ states by employing a microwave pulse and another blow-away pulse, the remaining $\sim \num{2e7}$ atoms are interrogated by three consecutive laser pulses forming a Mach-Zehnder interferometer.

The maximum pulse separation time is limited by the height of our interferometry tube and is $\SI{260}{ms}$. As the cloud falls, it eventually passes the detection region in which the relative population $P$ of the two ground states is determined via normalized fluorescence detection of $\sim \num{5e5}$ atoms within the sample.

\begin{figure}
    \centering
\begin{minipage}{0.48\textwidth}
\includegraphics[width=\textwidth]{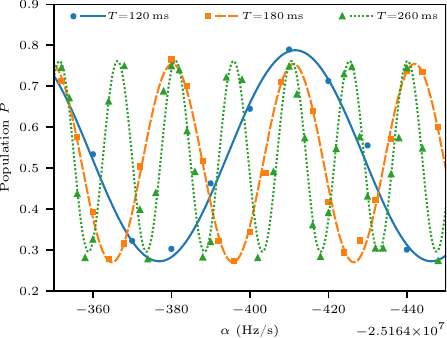}
\end{minipage}
    \caption{Interference fringes obtained by scanning the chirp rate $\alpha$ for three values of $T$. Lines are fits to eq. \eqref{eq:fringes_alpha_scan}.}
    \label{fig:alpha-scan}
\end{figure}

In order to obtain an interference fringe, this procedure is repeated  under variation of the two-photon detuning chirp rate $\alpha$ while the interferometry time $2T$ stays constant. Three interference fringes for different pulse separation times $T$ are shown in fig. \ref{fig:alpha-scan}. From these measurements, we obtain $C$ by a least-squares fit to

{}

\begin{equation}\label{eq:population}
    P(\alpha) = P_\text{mean} + \frac{1}{2} C \cos \Big((\alpha_0 - \alpha ) T^2\Big)
\end{equation}
where $P_\text{mean}$, $C$ and $\alpha_0$ are free parameters.

\begin{figure}
    \centering
\begin{minipage}{0.48\textwidth}
\includegraphics[width=\textwidth]{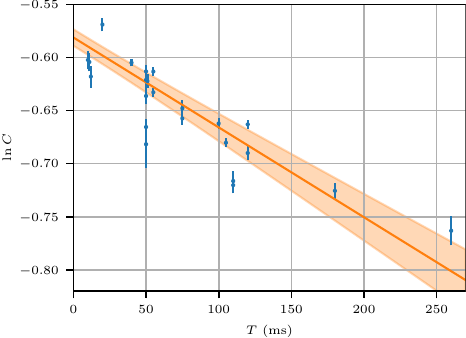}
\end{minipage}
\caption{Logarithm of the contrast obtained from 23 interference fringes as a function of the pulse separation time $T$. Error bars are $1\sigma$ fit uncertainties. The line is a weighted fit to eq. \ref{eq:contrastfit}. The shaded area is the 1$\sigma$ confidence interval.}
    \label{fig:result}
\end{figure}

As a proof-of-principle demonstration of the method presented in this paper, we analyze data that were collected during a previous gravity survey campaign \cite{Freier2016}. While the data has not specifically been recorded for this purpose, the data span a wide range of available values of $T$ and were recorded back-to-back.
Fig. \ref{fig:result} shows the contrast $C$ obtained from 23 interference fringes as a function of $T$.
The line is the result of a weighted least-squares fit of the contrast $C$ obtained from fitting eq. \eqref{eq:population} to
\begin{equation}\label{eq:contrastfit}
    \ln C(T) = \ln C_0 - 2\lambda_\text{CSL} N^2 T,
\end{equation}
where $\ln C_0$ and $\lambda_\text{CSL}$ are free parameters.
The loss of contrast $C_0$ independent of $T$ is due to the uncertainty of $\Delta\phi_0$ in eq. \eqref{eq:coherence}.

\begin{figure}
    \centering
\begin{minipage}{0.48\textwidth}
\includegraphics[width=\textwidth]{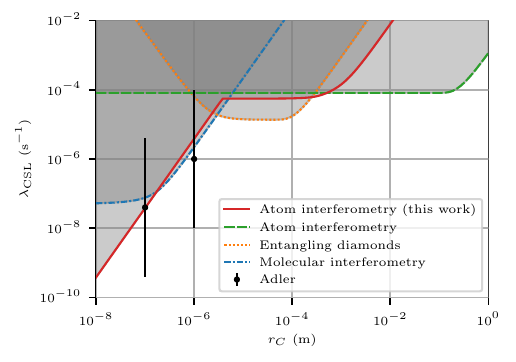}
\end{minipage}
\caption{Exclusion plot for the CSL parameters comparing different interferometric methods as compiled in Ref. \cite{Carlesso2019}: atom interferometry \cite{Kovachy2015} (green dashed and red solid line), interferometry with entangled diamonds \cite{Lee2011, Belli2016} (orange dotted line) and molecular interferometry \cite{Nimmrichter2011, Fein2019, Toros2017} (blue dash-dotted line). Shaded areas indicate the excluded regions and Adler's theoretical predictions \cite{Adler2007} are shown for comparison. Note that stronger bounds on the CSL parameters can be established by non-interferometric experiments not shown in this figure \cite{carlesso2022present}.}
    \label{fig:exclusion}
\end{figure}

 To give a conservative estimate, we neglect other mechanisms for loss of contrast inherent in the experiment and attribute them to a potential CSL effect. The CSL parameters excluded by our analysis are presented in fig.~\ref{fig:exclusion} alongside the results from other interferometric measurements.

In fig.~\ref{fig:exclusion} we evaluated our data explicitly including the term dependent on $r_C$ in eq. (\ref{eq:phivariance}) in order to get the correct bound for larger values of $r_C$.
For the approximated eq. (\ref{eq:contrastfit}), we obtain an upper bound on the collapse rate $\lambda_\text{CSL} \leq \SI{5.6(7)e-5}{\per\second}$ where the bound on $\lambda_\text{CSL}$ is not dependent on $r_C$.

As discussed in section \ref{sec:theory}, we can obtain a stronger bound for small $r_C$ by requiring that the two wave packets used in the interferometer have a relevant overlap at time $2T$. For our experimental parameters, this leads to a bound of $\lambda_{CSL}/r_{C}^{2} \leq \SI{3.9e6}{\per\meter\squared\per\second}$. This bound becomes stronger than $\lambda_\text{CSL} \leq \SI{5.6(7)e-5}{\per\second}$ for $r_C \leq \SI{3.8e-6}{m}$ and is thus reported as the diagonal section of the red solid line in fig. \ref{fig:exclusion}.

\section{Summary \& Outlook}

Using the path integral formalism, we derived the variance of the phase difference occurring in a Mach-Zehnder atom interferometer due to the nonlinear and stochastic additions of CSL to the Schr\"{o}dinger equation.
This yields an exponential decrease of the contrast proportional to the interrogation time $T$ and the square of the number of nucleons $N^2$.
We measured the contrast for $T$ ranging from  $\SI{11}{ms}$ up to $\SI{260}{ms}$ and found bounds for the CSL parameters reported in fig. \ref{fig:exclusion}.

This result could further be improved by characterizing other mechanisms of contrast loss.
We identify two major sources for decreased contrast in our experiment from numerical simulations: (i)~the finite size and temperature of the atomic sample and its convolution with the Gaussian intensity profile of the laser beams leading to varying Rabi frequencies and, thus, reduced transfer efficiencies of the laser pulses; (ii)~the finite temperature of the atomic sample along the laser beams giving rise to Doppler shifts, again resulting in reduced transfer efficiencies. However, these effects have not been studied experimentally for our setup, yet.

Using modern tools of atom interferometry, the bounds on $\lambda_\text{CSL}$ could be further lowered.
By employing high momentum transfer beam splitters \cite{Gebbe2021} CSL could be tested for larger $r_C$ using our method. Further improvements are within reach using ultracold atomic sources and squeezing techniques \cite{Schrinski2020} and the mechanisms of contrast loss could be studied at extended interferometry times $T$ such as in large atomic fountains \cite{Kovachy2015}, optical lattices \cite{Xu2019},  microgravity experiments \cite{Muntinga2013}, or in space~\cite{Lachmann2021, Aveline2020}.

\begin{acknowledgements}
We thank Angelo Bassi, Holger M\"{u}ller, Matteo Carlesso and Francesco Intravaia for helpful comments.

This work is supported by the German Space Agency DLR with funds provided by the Federal Ministry of Economics and Technology (BMWi) under grant numbers DLR 50WM1953, 50WP1702, 50WP2102, 50WM2052 and 50WP1432. S.D. acknowledges support from the Fetzer Franklin Fund and INFN.
\end{acknowledgements}

\section*{Author contributions}
\textbf{S. Vowe}: Conceptualization (lead); Data curation (supporting); Formal analysis (supporting);  Methodology (supporting); Visualization (supporting); Writing - original draft (equal); Writing - review \& editing (supporting).
\textbf{S. Donadi}: Conceptualization (supporting); Methodology (lead); Visualization (supporting); Writing - original draft (equal); Writing - review \& editing (supporting)
\textbf{V. Schkolnik}:  Data curation (lead); Writing - original draft (supporting); Writing - review \& editing (supporting).
\textbf{A. Peters}: Resources (equal); Writing - original draft (supporting)
\textbf{B. Leykauf}: Data curation (supporting); Formal analysis (lead); Project administration (equal); Visualization (lead); Writing - original draft (supporting); Writing - review \& editing (lead).
\textbf{M. Krutzik}: Conceptualization (supporting); Project administration (equal); Resources (equal); Writing - original draft (supporting); Writing - review \& editing (supporting).

\onecolumngrid
\appendix
\section{Computation of eq. (\ref{eq:phivariance}) of the main text}\label{app:a}
The evolution of $z(t)$ and $\Dot{z}(t)$ in the free falling frame along the paths $ABC$ and $ADC$ is (see fig. \ref{fig:ai-sketch} of the main text) 
\begin{align*}\label{paths}
    &\text{AB:}\quad z_{AB}(t) = v_2 t, \quad \Dot{z}_{AB} = v_2 \\
    &\underbrace{\text{AD:}\quad z_{AD}(t) = v_1 t, \quad \Dot{z}_{AD} = v_1}_{t\in[0,T] }  \\
    &\text{BC:}\quad z_{BC}(t) = v_2 T + v_1 (t-T), \quad \Dot{z}_{BC} = v_1 \\
    &\underbrace{\text{DC:}\quad z_{DC}(t) = v_1 T + v_2 (t-T), \quad \Dot{z}_{DC} = v_2 }_{t\in[T,2T]}\nonumber
\end{align*}{}

The terms in eqs. (4) and (5) of the main text can be evaluated by calculating the integrals along these paths. For example, the integral along the path ABC is given by

\begin{equation}
    \int_{ABC}V_\text{CSL}(z(t), t)\mathrm{d}t  = \int_{0}^T V_\text{CSL}(z_{AB}(t), t )\mathrm{d}t +\int_{T}^{2T}V_\text{CSL}(z_{BC}(t), t )\mathrm{d}t
\end{equation}
and similarly for the other one.

We now derive the result in eq. (5):

\begin{equation}\label{eq:phivar}
\begin{split}
    \mathbb{E}[\Delta\phi^2_\text{CSL}] &=
    \mathbb{E}\bigg[\frac{1}{\hbar^2} \ointop_{ABCDA}\ointop_{ABCDA}
    V_\text{CSL}(z(t), t)V_\text{CSL}(z(t'), t')\mathrm{d}t\mathrm{d}t'\bigg]
    \\
    &=
    \mathbb{E}\bigg[\frac{1}{\hbar^2}  
    \bigg(
    \intop_{ABC}V_\text{CSL}(z(t), t)-\intop_{ADC}V_\text{CSL}(z(t), t)
    \bigg)
    \bigg(
    \intop_{ABC}V_\text{CSL}(z(t'), t')-\intop_{ADC}V_\text{CSL}(z(t'), t')
    \bigg)
    \mathrm{d}t
    \mathrm{d}t'
    \bigg]
    \\
    &=
    \phi^{UU}-\phi^{UD}-\phi^{DU}+\phi^{DD},
\end{split}
\end{equation}
where the superscript $U$ refers to the upper path ABC and $D$ to the lower path ADC (see fig. 1 in the main text). In eq. (\ref{eq:phivar}), each of the four terms introduced in the last line has the form (here and below $X=U,D$ and $Y=U,D$)
\begin{equation}
\phi^{XY}:=\mathbb{E}\left[\frac{1}{\hbar^{2}}\int_{0}^{2T}\mathrm{d}tV_\text{CSL}^{X}(t)\int_{0}^{2T}\mathrm{d}t'V_\text{CSL}^{Y}(t')\right]\label{phi_xy}
\end{equation}
and 
\begin{equation}
V_{CSL}^{X}(t)=-\frac{\hbar\sqrt{\lambda_\text{CSL}}N}{(\sqrt{\pi}r_{C})^{3/2}}\int \mathrm{d}\boldsymbol{s}\,w(\boldsymbol{s},t)e^{-\frac{(\boldsymbol{s}-\boldsymbol{q}^{X}(t))^{2}}{2r_{C}^{2}}}\label{V_csl-1-1}
\end{equation}
with $\boldsymbol{q}^{U(D)}(t)$ the trajectory along the points ABC(ADC). Then eq. (\ref{phi_xy}) becomes
\begin{equation}
\begin{split}
\phi^{XY}&=\frac{\lambda_\text{CSL} N^{2}}{(\sqrt{\pi}r_{C})^{3}}\int_{0}^{2T}\mathrm{d}t\int_{0}^{2T}\mathrm{d}t'\int d\boldsymbol{s}\int \mathrm{d}\boldsymbol{s}'e^{-\frac{(\boldsymbol{s}-\boldsymbol{q}^{X}(t))^{2}}{2r_{C}^{2}}}e^{-\frac{(\boldsymbol{s}'-\boldsymbol{q}^{Y}(t'))^{2}}{2r_{C}^{2}}}\mathbb{E}\left[w(\boldsymbol{s},t)w(\boldsymbol{s}',t')\right]
\\
&=\frac{\lambda_\text{CSL} N^{2}}{(\sqrt{\pi}r_{C})^{3}}\int_{0}^{2T}\mathrm{d}t\int \mathrm{d}\boldsymbol{s}e^{-\frac{(\boldsymbol{s}-\boldsymbol{q}^{X}(t))^{2}}{2r_{C}^{2}}}e^{-\frac{(\boldsymbol{s}-\boldsymbol{q}^{Y}(t))^{2}}{2r_{C}^{2}}}\\
&=\lambda_\text{CSL} N^{2}\int_{0}^{2T}\mathrm{d}t\,e^{-\frac{\left(\boldsymbol{q}^{X}(t)-\boldsymbol{q}^{Y}(t)\right)^{2}}{4r_{C}^{2}}}.
\end{split}
\end{equation}
We see immediately that 
\begin{equation}\label{UU}
\phi^{UU}=\phi^{DD}=2\lambda_\text{CSL} N^{2}T,
\end{equation}
while
\begin{equation}\label{UD}
\begin{split}
\phi^{UD}=\phi^{DU}&=\lambda_\text{CSL} N^{2}\int_{0}^{2T}\mathrm{d}t\,e^{-\frac{\left(z^{U}(t)-z^{D}(t)\right)^{2}}{4r_{C}^{2}}}
\\
&=\lambda_\text{CSL} N^{2}\left(\int_{0}^{T}\mathrm{d}t\,e^{-\frac{\left(z_{AB}(t)-z_{AD}(t)\right)^{2}}{4r_{C}^{2}}}+\int_{T}^{2T}\mathrm{d}t\,e^{-\frac{\left(z_{BC}(t)-z_{DC}(t)\right)^{2}}{4r_{C}^{2}}}\right)
\\
&=\lambda_\text{CSL} N^{2}\left(\int_{0}^{T}\mathrm{d}t\,e^{-\frac{\left(v_{2}-v_{1}\right)^{2}t^{2}}{4r_{C}^{2}}}+\int_{T}^{2T}\mathrm{d}t\,e^{-\frac{(v_{2}-v_{1})^{2}(2T-t)^{2}}{4r_{C}^{2}}}\right)
\\
&=\lambda_\text{CSL} N^{2}\frac{\sqrt{\pi}\text{erf}\left[\left(\frac{v_{2}-v_{1}}{2r_{C}}\right)T\right]}{\left(\frac{v_{2}-v_{1}}{2r_{C}}\right)},
\end{split}
\end{equation}
where we used\footnote{The paths considered here are those seen by an observer in the free falling frame. However, the result would be the same for an observer in the laboratory frame: this is because the effect of CSL on the phase shift depends only on the difference between the two paths, which is the same in both frames.
}
\begin{equation}
z_{AB}(t)-z_{AD}(t)=(v_{2}-v_{1})t,\;\;\;\;\;\;z_{BC}(t)-z_{DC}(t)=(v_{2}-v_{1})(2T-t).
\end{equation}
and the fact that $x^{U}(t)=x^{D}(t)$ and $y^{U}(t)=y^{D}(t)$ as the paths differ only along the $z$ direction.
Inserting the results from eqs. (\ref{UU}) and (\ref{UD}) into eq. (\ref{eq:phivar}) one obtains the result in eq. (\ref{eq:phivariance}) in the main text.

\section{An analysis using wave packets}\label{app:b}

In this appendix, we discuss CSL effects on atom interferometry performing a more detailed analysis which accounts for the finite extent of the wave packets. As we will see, this will not change the decay of the coherences between the wave packets, confirming the results from the calculation in the main text. However, we will also see that a new effect arises, which is relevant for small $r_C$: by requiring that the two wave packets overlap at the end of the interferometer (i.e. at time $t=2T$), we can set another bound on the CSL parameters.

We start by modelling the state of each atom, after the action of the first $\pi/2$-pulse,
as a superposition of two Gaussian wave packets with width $\sigma$: 
\begin{equation}
|\psi(0)\rangle=\frac{|\psi_{1}\rangle+|\psi_{2}\rangle}{\sqrt{2}}\label{psi0}
\end{equation}
where
\begin{equation}
\psi_{j}(\boldsymbol{x})=\frac{1}{\left(\sqrt{2\pi}\sigma\right)^{\frac{3}{2}}}e^{-\frac{\boldsymbol{x}^{2}}{4\sigma^{2}}+ik_{j}z}\;\;\;\;\;\;\textrm{with}\;\;\;\;\;\;k_{1}\neq k_{2}\text{.}\label{psij}
\end{equation}
The corresponding statistical operator is 
\begin{equation}
\rho(0)=\frac{1}{2}\sum_{i,j=1}^{2}|\psi_{i}\rangle\langle\psi_{j}|.\label{ro0}
\end{equation}
Given this initial state, the free evolution of the statistical operator from 0 to $T$ can be computed using the relation~\cite{GRW1986}
\begin{equation}
\rho_{ij}^\text{CSL}(\boldsymbol{x},\boldsymbol{y},T)=\frac{1}{(2\pi)^{3}}\int \mathrm{d}\boldsymbol{k}\int \mathrm{d}\boldsymbol{w}e^{-i\boldsymbol{k}\cdot\boldsymbol{w}}F^\text{CSL}(\boldsymbol{k},\boldsymbol{x}-\boldsymbol{y},T)\rho_{ij}^{QM}(\boldsymbol{x}+\boldsymbol{w},\boldsymbol{y}+\boldsymbol{w},T)\label{ev_csl}
\end{equation}
where 
\begin{equation}
F^\text{CSL}(\boldsymbol{k},\boldsymbol{x}-\boldsymbol{y},t)=\exp\left[-\lambda_\text{CSL}\frac{m^{2}}{m_{0}^{2}}t\left(1-\frac{1}{t}\int_{0}^{t}\mathrm{d}\tau\,e^{-\frac{\left(\boldsymbol{x}-\boldsymbol{y}-\frac{\hbar\boldsymbol{k}t}{m}\right)^{2}}{4r_{C}^{2}}}\right)\right]\label{F}
\end{equation}
and $\rho_{ij}^{QM}$ is the statistical operator evolved according to the free Schr\"odinger equation
\begin{equation}
\rho_{ij}^\text{QM}\left(\boldsymbol{x},\boldsymbol{y},T\right)=e^{i(\boldsymbol{k}_{i}\cdot\boldsymbol{x}-\boldsymbol{k}_{j}\cdot\boldsymbol{y})}\psi_{i}(\boldsymbol{x})\psi_{j}^{*}(\boldsymbol{y})\label{ro_qm}
\end{equation}
where
\begin{equation}
\psi_{j}(\boldsymbol{x})=\frac{1}{\left[\frac{\sqrt{2\pi}}{\sigma}\sigma_{T}^{2}\right]^{\frac{3}{2}}}e^{-\frac{(\boldsymbol{x}-\boldsymbol{x}_{j})^{2}}{4\sigma_{T}^{2}}},\label{psi ev}
\end{equation}
with $\boldsymbol{x}_{j}:=\frac{\hbar T}{m}\boldsymbol{k}_{j}$ and
$\sigma_{T}=\sigma\sqrt{1+\frac{i\hbar T}{2m\sigma^{2}}}$. 

Starting from eq. (\ref{ev_csl}), we first perform the integration in $\boldsymbol{w}$, that gives:
\begin{equation}\label{I_ij}
\begin{split}
I_{ij}(\boldsymbol{x},\boldsymbol{y},T)&:=\int \mathrm{d}\boldsymbol{w}e^{-i\boldsymbol{k}\cdot\boldsymbol{w}}\rho_{ij}^\text{QM}(\boldsymbol{x}+\boldsymbol{w},\boldsymbol{y}+\boldsymbol{w},T) \\
&=\!
\left(\frac{\sigma}{\sigma_{T}}\right)^{3}\!\!\exp{\frac{i}{2}[\boldsymbol{k}\!\cdot\!(\boldsymbol{x}-\boldsymbol{x}_{i}-\boldsymbol{x}_{j}+\boldsymbol{y})+\boldsymbol{k}_{i}\!\cdot\!(\boldsymbol{x}+\boldsymbol{x}_{i}+\boldsymbol{x}_{j}-\boldsymbol{y})+\boldsymbol{k}_{j}\!\cdot\!(\boldsymbol{x}-\boldsymbol{x}_{i}-\boldsymbol{x}_{j}-\boldsymbol{y})]}\times\\
&\times e^{-\frac{\sigma_{T}^{2}}{2}(\boldsymbol{k}-\boldsymbol{k}_{i}+\boldsymbol{k}_{j})^{2}}e^{-\frac{(\boldsymbol{x}-\boldsymbol{y}-\boldsymbol{x}_{i}+\boldsymbol{x}_{j})^{2}}{8\sigma_{T}^{2}}}
\end{split}
\end{equation}
and eq. (\ref{ev_csl}) becomes
\begin{equation}
\rho_{ij}^\text{CSL}(\boldsymbol{x},\boldsymbol{y},T)=\frac{1}{(2\pi)^{3}}\int \mathrm{d}\boldsymbol{k}F^\text{CSL}(\boldsymbol{k},\boldsymbol{x}-\boldsymbol{y},T)I_{ij}(\boldsymbol{x},\boldsymbol{y},T).\label{roij}
\end{equation}
This equation will be the starting point for the following analysis.
First, we will study the dynamics of the off-diagonal terms ($i\neq j$) and show that they are not affected by the finite size of the wave packets for all values of $r_C$ (\textbf{A}); then we show how CSL diffusion affects the diagonal elements, i.e. the dynamics of each wave packet (\textbf{B}). By requiring that the wave packets have a relevant overlap at time $2T$, we find a bound on the ratio $\lambda/r_C^2$.

\subsection{Study of the off-diagonal terms}

We study the term $\rho_{12}^\text{CSL}$. Then eq. (\ref{roij}) becomes
\begin{equation}\label{rho1212}
\rho_{12}^\text{CSL}(\boldsymbol{x},\boldsymbol{y},T)=\frac{1}{(2\pi)^{3}}\int \mathrm{d}\boldsymbol{k}F^\text{CSL}(\boldsymbol{k},\boldsymbol{x}-\boldsymbol{y},T)I_{12}(\boldsymbol{x},\boldsymbol{y},T).
\end{equation}
The calculation for $\rho_{21}^\text{CSL}$ is identical with the simple replacement $\boldsymbol{k}_{1}\leftrightarrow\boldsymbol{k}_{2}$.
Looking at the terms in the last line of eq. (\ref{I_ij}), one can see that the only relevant
contributions to the integral are in the range where
\begin{equation}
|\boldsymbol{k}|\simeq|\boldsymbol{k}_{1}-\boldsymbol{k}_{2}|\pm\ell_{T}^{-1}
\quad\text{and}\quad
|\boldsymbol{x}-\boldsymbol{y}|\simeq|\boldsymbol{x}_{1}-\boldsymbol{x}_{2}|\pm\ell_{T}\label{xy_12}
\end{equation}
with $\ell_{T}=\sigma\sqrt{\left(1+\frac{\hbar^{2}T^{2}}{4m^{2}\sigma^{4}}\right)}$
($\sigma_{T}$ cannot be taken directly as the width of the Gaussian,
being a complex quantity). For the setup considered here, we have
\[
|\boldsymbol{k}_{1}-\boldsymbol{k}_{2}|\simeq1.6\times10^{7}\;\textrm{m}^{-1},\;\;\;
|\boldsymbol{x}_{1}-\boldsymbol{x}_{2}|\simeq 3\times10^{-3}\;\textrm{m}\;\;\text{and}\;\;
\ell_{T}\simeq7\times10^{-5}\;\textrm{m},
\]

(to compute $\ell_T$ we took $\sigma=10^{-6}$ m and $T=190$ ms). Hence, $|\boldsymbol{x}_{1}-\boldsymbol{x}_{2}|\gg\ell_{T}$ and $|\boldsymbol{k}_{1}-\boldsymbol{k}_{2}|\gg\ell_{T}^{-1}$.
This confirms that if the wave packet size is much smaller than the distance between the two, the details about their extent are irrelevant when studying the coherences.
Thus, we can approximate the factor accounting for CSL effects in eq. (\ref{F}) as
\begin{equation}
F^\text{CSL}(\boldsymbol{k},\boldsymbol{x}-\boldsymbol{y},t)\simeq\exp\left[-\lambda_\text{CSL}\frac{m^{2}}{m_{0}^{2}}t\left(1-\frac{1}{t}\int_{0}^{t}\mathrm{d}\tau\,e^{-\frac{\left(\boldsymbol{x}_{1}-\boldsymbol{x}_{2}-\frac{\hbar(\boldsymbol{k}_{1}-\boldsymbol{k}_{2})\tau}{m}\right)^{2}}{4r_{C}^{2}}}\right)\right]\label{F12}.
\end{equation}
i.e. the CSL factor becomes independent of $\boldsymbol{k}$ and $\boldsymbol{x}-\boldsymbol{y}$ and can be taken out of the integral in eq. (\ref{rho1212}). This implies
\begin{align}
\rho_{12}^\text{CSL}(\boldsymbol{x},\boldsymbol{y},T)&\simeq\exp\left[-\lambda_\text{CSL}\frac{m^{2}}{m_{0}^{2}}T\left(1-\frac{1}{T}\int_{0}^{T}\mathrm{d}\tau\,e^{-\frac{\left(\boldsymbol{x}_{1}-\boldsymbol{x}_{2}-\frac{\hbar(\boldsymbol{k}_{1}-\boldsymbol{k}_{2})\tau}{m}\right)^{2}}{4r_{C}^{2}}}\right)\right]\rho_{12}^\text{QM}(\boldsymbol{x},\boldsymbol{y},T)\nonumber\\
&=\exp\left[-\lambda_\text{CSL}\frac{m^{2}}{m_{0}^{2}}T\left(1-\frac{1}{T}\int_{0}^{T}\mathrm{d}\tau\,e^{-\frac{\hbar^{2}}{4r_{C}^{2}m^{2}}(\boldsymbol{k}_{1}-\boldsymbol{k}_{2})^{2}\tau^{2}}\right)\right]\rho_{12}^\text{QM}(\boldsymbol{x},\boldsymbol{y},T)
\end{align}
where we used $\boldsymbol{x}_{j}:=\frac{\hbar T}{m}\boldsymbol{k}_{j}$ and performed the change of variable $T-\tau\rightarrow\tau$.\\
We have shown that CSL effects on the off diagonal elements are independent from the wave functions spatial extent and will now we take a closer look at two relevant regimes  of $r_C$:
\begin{itemize}
\item $r_{C}\gg|\boldsymbol{x}_{1}-\boldsymbol{x}_{2}|$: In this case,
since $\frac{\hbar|\boldsymbol{k}_{1}-\boldsymbol{k}_{2}|\tau}{m}\leq|\boldsymbol{x}_{1}-\boldsymbol{x}_{2}|$
(with the equivalence when $\tau=T)$, we can expand the exponential
and obtain
\begin{equation}\label{ro12large}
\rho_{12}^\text{CSL}(\boldsymbol{x},\boldsymbol{y},T)\simeq\exp\left[-\frac{\lambda_\text{CSL}\hbar^{2}}{12r_{C}^{2}m_{0}^{2}}(\boldsymbol{k}_{1}-\boldsymbol{k}_{2})^{2}T^{3}\right]\rho_{12}^\text{QM}(\boldsymbol{x},\boldsymbol{y},T).
\end{equation}

\item $r_{C}\ll|\boldsymbol{x}_{1}-\boldsymbol{x}_{2}|$: Here, one gets 
\begin{equation}\label{ro12small}
\rho_{12}^\text{CSL}(\boldsymbol{x},\boldsymbol{y},T)\simeq\exp\left[-\lambda_\text{CSL}\frac{m^{2}}{m_{0}^{2}}T\right]\rho_{12}^\text{QM}(\boldsymbol{x},\boldsymbol{y},T).
\end{equation}
\end{itemize}
Eqs. (\ref{ro12large}), (\ref{ro12small}) describe the loss of coherences
due to CSL from time 0 to time $T$. The evolution from $T$ to $2T$,
after the $\pi$ pulse, it is precisely the same, just starting with
the two wave packets spatially separated and then converging to the
same point. This implies that the final damping factor will be the
same as in eqs. (\ref{ro12large}), (\ref{ro12small}) only with the
exponents doubled. These damping factors are in agreement with the result in the main text.

\subsection{Study of the diagonal terms}

We now consider the evolution of the diagonal terms which are of the form 
\begin{equation}
\rho_{ii}^\text{CSL}(\boldsymbol{x},\boldsymbol{y},T)=\frac{1}{(2\pi)^{3}}\int \mathrm{d}\boldsymbol{k}F^\text{CSL}(\boldsymbol{k},\boldsymbol{x}-\boldsymbol{y},T)I_{ii}(\boldsymbol{x},\boldsymbol{y},T)\label{roii}
\end{equation}
with $i=1,2$ and 
\begin{equation}
I_{ii}(\boldsymbol{x},\boldsymbol{y},T)=\left(\frac{\sigma}{\sigma_{T}}\right)^{3}\exp\left\{ \frac{i}{2}\left[\boldsymbol{k}\cdot(\boldsymbol{x}-2\boldsymbol{x}_{i}+\boldsymbol{y})+2\boldsymbol{k}_{i}\cdot(\boldsymbol{x}-\boldsymbol{y})\right]\right\} e^{-\frac{\sigma_{T}^{2}}{2}\boldsymbol{k}^{2}}e^{-\frac{(\boldsymbol{x}-\boldsymbol{y})^{2}}{8\sigma_{T}^{2}}}\label{Iii}
\end{equation}
In this case, the two Gaussian in eq. (\ref{Iii}) imply that the relevant ranges
for $\boldsymbol{k}$ and $\boldsymbol{x}-\boldsymbol{y}$ are 
\begin{gather}
|\boldsymbol{k}|\leq\pm\ell_{T}^{-1}\\
|\boldsymbol{x}-\boldsymbol{y}|\leq \pm\ell_{T}.
\end{gather}

As before, we consider two regimes:

\begin{itemize}
\item $r_{C}\gg\ell_{T}$: Here, the second term in the exponent of the CSL factor $F^\text{CSL}$ in eq. (\ref{F}) is always very close to 1, hence, to the lowest order CSL effects are negligible.
This is expected since by studying the diagonal elements we are focusing on the evolution of each single wave packet: when the size of a wave packet $\ell_T$ is much smaller than $r_C$, it is well known that CSL effects become negligible. 

\item $r_{C}\leq\ell_{T}$: In this regime the diagonal elements are affected 
by the CSL induced collapse. This does not come as a surprise since, heuristically, $r_{C}$ gives the spatial resolution of the collapse. 
The analysis of this regime is non-trivial because there are no approximations valid for all times to simplify $F^\text{CSL}$ in eq. (\ref{F}).
However, this regime is not relevant for our analysis since for such small values of $r_C$ we can set much stronger bounds by requiring that the two wave packets have to overlap at the same region in space at time $2T$.
\end{itemize}

In fact, the CSL-induced noise responsible for the collapse also implies (together with the collapse of spatial superposition) a diffusion (quite similar to Brownian motion) that induces an increase of the average energy of the system. This heating effect for a mass $m$ is given by CSL heating rate is \cite{bassi2003dynamical}
\begin{equation}\label{heating}
\langle H_{t}\rangle=\langle H_{0}\rangle+\frac{3m\hbar^{2}\lambda_\text{CSL}}{4m_{0}^{2}r_{C}^{2}}t.
\end{equation}

Corresponding to this heating, there is an increase in the position variance, which, focusing just in the $z$ direction where interferometry is performed, is given by
\begin{equation}\label{eq:std}
    \langle \Delta z^2\rangle_{t}=\langle \Delta z^2\rangle^\text{QM}_{t}+\langle \Delta z^2\rangle^\text{CSL}_{t}
\end{equation}
where $\Delta z^{2}:=z^{2}-\langle z\rangle^{2}$, $\langle \Delta z^2\rangle^\text{QM}_{t}$ is the standard quantum mechanical spread and
\begin{equation}
    \langle \Delta z^2\rangle^\text{CSL}_{t}=\frac{\lambda_\text{CSL}\hbar^{2}}{6m_{0}^{2}r_{C}^{2}}t^{3}.
\end{equation}
The last term accounts for the diffusion in space induced by CSL. It should be noted that eq. (\ref{eq:std}) describes the spread of a statistical ensemble of atoms, not the spread of the individual atoms' wave functions. Furthermore note that this increase of variance due to CSL is not in contradiction with the fact that the model collapses in position. The effect of CSL is indeed to shrink the wave packet, compared to what one would obtain just with only the Schr\"odinger evolution \cite{bassi2003dynamical}. However, the location where the wave packet will collapse is not always the same (otherwise there would be a clear violation of the Born rule). Overall this leads to an increase in the position variance, $\langle \Delta z^2\rangle^\text{CSL}_{t}$, which quantifies the extension of the space region where the wave packet is randomly diffused. For a detailed and quantitative analysis of this point with a simpler collapse model where all the dynamics can be solved exactly, see \cite{bassi2005collapse}.

In order to see interferometric effects, a necessary condition is that the two wave packets recombine after the last $\pi/2$ pulse. This is not possible if the diffusion in position induced by the CSL model $\sqrt{\langle \Delta z^2\rangle^\text{CSL}_{t}}$ is much larger than the wave packet size $\ell_{2T}=\sqrt{\langle \Delta z^2\rangle^\text{QM}_{t}}$ at the final time $2T$. Hence, to be conservative we require $\sqrt{\langle \Delta z^2\rangle^\text{CSL}_{t}}\leq 0.1\times\ell_{2T}$, which leads to
\begin{equation}\label{bound_diag}
\frac{\lambda_\text{CSL}}{r_{C}^{2}}\leq\frac{0.01 \cdot \ell_{2T}^{2}6m_{0}^{2}}{\hbar^{2}(2T)^{3}}\simeq 3.9 \times10^{6}\;\textrm{m}^{-2}\textrm{s}^{-1},
\end{equation}
where we took according to the experiment $2T \simeq \SI{520}{ms}$ and $\ell_{2T}=\sigma  \sqrt{1+\frac{ \hbar^2 (2T)^2 }{4 m^2 \sigma ^4}}\simeq \SI{1.9e-4}{m}$ when taking the initial wave packet spread as $\sigma=\SI{e-6}{m}$ and $m=\SI{1.44e-25}{kg}$ the mass of a $^{87}$Rb atom. Note that inequality (\ref{bound_diag}) states a necessary condition to observe interference in our experiment. Thus, we can exclude the part of the parameter space where eq. (\ref{bound_diag}) is \emph{not} fulfilled.  Furthermore note that, contrary to typical interference experiments where the reduction of the fringes visibility is due to the accumulation of different phases by each wave packet of the superposition, here the loss of contrast is due to the possibly small overlap of the two wave packets at final time $2T$. 

The bound in eq. (\ref{bound_diag}) corresponds to the diagonal bound reported in Figure 5 in the main text. In particular, when $r_{C}\lesssim \SI{e-6}{m}$, this bound is stronger than the one found due to the interferometric effects described in section \ref{app:a} of the appendix. We note that this bound is conservative since we took as a size of the wave packet $\ell_{2T}$ which accounts only for the Schr\"odinger evolution; if one includes also the shrinking of the wave packet due to collapse, the bound may become even stronger. 

\bibliography{main}

\end{document}